\documentclass[conference]{IEEEtran}
\IEEEoverridecommandlockouts
\usepackage{cite}
\usepackage{amsmath,amssymb,amsfonts}
\usepackage{graphicx}
\usepackage{textcomp}
\usepackage{xcolor}
\usepackage{bbm}
\usepackage{algorithm}
\usepackage{algorithmic}
\usepackage{setspace}
\usepackage{subfigure}
\usepackage{xcolor}
\usepackage{url}
\usepackage[T1]{fontenc}

\newcommand{\Mod}[1]{\ (\mathrm{mod}\ #1)}
\DeclareMathOperator*{\Ex}{\mathbb{E}}
\usepackage{dsfont} 

\usepackage{mathtools}

\definecolor{lred}{rgb}{0.9, 0.2, 0.2}

\usepackage{bm}
\graphicspath{{./figures/}}
\usepackage{booktabs}
\usepackage{outlines}

\newcommand{\dcul}{\mbox{DC-ULCB}}
\DeclareGraphicsExtensions{.pdf,.jpeg,.png,.eps} %

\usepackage{amsthm}
\def\BibTeX{{\rm B\kern-.05em{\sc i\kern-.025em b}\kern-.08em
    T\kern-.1667em\lower.7ex\hbox{E}\kern-.125emX}}
\newtheorem{myTheo}{Theorem}
\newtheorem{lemma}{Lemma}

\begin{document}

\title{
Fair Distributed Cooperative Bandit Learning on Networks for Intelligent Internet of Things Systems\
\emph{Technical Report}
}

\author{
Ziqun~Chen\textsuperscript{*}, Kechao~Cai\textsuperscript{*},
Jinbei~Zhang\textsuperscript{*},
Zhigang~Yu\textsuperscript{$\dagger$}\\
\textsuperscript{*}{School of Electronics and Communication Engineering,
Sun Yat-sen University, Shenzhen, China} \\
\textsuperscript{$\dagger$}{China Academy of Electronics and Information Technology,
Beijing, China} \\
chenzq35@mail2.sysu.edu.cn, \{caikch3,zhjinbei\}@mail.sysu.edu.cn, yzg11@tsinghua.org.cn}

\maketitle

\begin{abstract}
In intelligent Internet of Things (IoT) systems, edge servers within a network exchange information with their neighbors and collect data from sensors to complete delivered tasks.
In this paper, we propose a multiplayer multi-armed bandit model for intelligent IoT systems to facilitate data collection and incorporate fairness considerations. 
In our model, we establish an effective communication protocol that helps servers cooperate with their neighbors.
Then we design a distributed cooperative bandit algorithm, \dcul{}, enabling servers to collaboratively select sensors to maximize data rates while maintaining fairness in their choices.
We conduct an analysis of the reward regret and fairness regret of \dcul{}, and
prove that both regrets have logarithmic instance-dependent upper bounds. 
Additionally, through extensive simulations, we validate that \dcul{}
outperforms existing algorithms in maximizing reward and ensuring fairness.
 
\end{abstract}



\section{Introduction}
\label{sec:introduction}

With the rapid development of wireless communication and edge computing, intelligent Internet of Things (IoT) systems come into focus in recent years~\cite{li2018learning}.
Due to the limited network transmission performance, the centralized server becomes inefficient for processing and analyzing huge volumes of data collected from sensors. 
Intelligent IoT systems deliver computing tasks to the multiple networked edge servers, which gather information from data-collecting sensors via wireless channels~\cite{luong2016data}.
Then edge servers explore data efficiently and complete delivered tasks through local computation and communication with each other~\cite{liu2022tutorial}.
To facilitate data collection, a distributed cooperative learning algorithm is required to maximize data rates from sensors to servers.

Due to the limited capacity of edge servers, each server can only collect data from a sensor at a time.
Therefore, it is favorable for servers to select the sensors with higher data rates. 
However, the channel state information between servers and sensors is usually unknown at the beginning.
A server has to estimate these parameters while selecting the sensors to maximize its data rates.
At this point, each server faces a dilemma: explore various sensors to
learn the channel information or exploit the historical observations to select the sensors with a high data rate in hindsight.
Moreover, it is important to ensure that every server receives a fair
allocation of sensors to prevent certain servers from monopolizing the
best sensors. This fair sensor selection requirement can greatly enhance the data rates for less fortunate servers.

The multiplayer multi-armed bandit (MMAB)~\cite{lattimore2020bandit,cai2022learning}
turns out to be an effective model for resolving the dilemma above.
Specifically, we view servers as players and data-collecting sensors as arms.
Each arm has an unknown reward distribution and the goal is to maximize
the cumulative reward (data rates) over all players based on their observations
and decision history in pulling the arms. 
In intelligent IoT systems, servers distributed at the network edge can exchange information with their neighbors through communication links.
Therefore, we demand an effective communication protocol that helps servers cooperate with their neighbors.
Moreover, collisions (servers selecting the same sensor) also become frequent as
servers all tend to select the sensors with higher rewards (data rates) if there is no coordination. 
The data transferring would be interrupted when a collision occurs and no data
can transmit successfully, corresponding to a loss of reward in the MMAB model. 
Thus, the servers would better select different sensors to avoid collisions and a
fairness issue arises: which server should use the best sensor with the highest data rate?
Therefore, it is challenging to design a distributed algorithm that both allows for 
limited server communications and ensures fair sensor selection so that all servers can experience the same data rates.



In our work, we propose an MMAB model for intelligent IoT systems where
each server can not only communicate with their neighbors in a network but also
receive a fair share of the reward.
We design a novel distributed cooperative bandit algorithm called \dcul{} for
servers.
Specifically, each server first runs an initialization scheme and incorporates a
running consensus procedure to estimate the expected reward of each sensor
cooperatively. 
Using these estimates, each server can then select the best sensor associated with its rank to avoid collisions. 
We prove that \dcul{} can maximize the cumulative reward of servers while
satisfying the fairness requirement. 

Our main contributions are listed as follows: (\romannumeral1) To our best
knowledge, we are the first to consider an MMAB model with both server
communications and fair sensor selection for intelligent IoT systems.
(\romannumeral2) We design an algorithm \dcul{} for each server to select the
sensors and maximize their rewards cooperatively while guaranteeing fair
sensor selections by incorporating a running consensus procedure.
(\romannumeral3) We prove that \dcul{} achieves instance-dependent logarithmic
reward regret upper bound and fairness regret upper bound.
We conduct extensive simulations to demonstrate the effectiveness of \dcul{}. 
The simulation results show that our algorithm significantly outperforms the
existing algorithms.

The rest of this paper is organized as follows.
We give a brief literature overview in Sec.~\ref{sec:mmab:related-work}.
We introduce our multi-armed bandit model in Sec.~\ref{sec:syst-model-perf}.
We describe the details of our \dcul{} algorithm and provide a theoretical
analysis of \dcul{} in Sec.~\ref{sec-distributed-policy}.
We present our simulation results in Sec.~\ref{sec-numerical-experiments}. 
We conclude the paper in Sec.~\ref{sec-conclusion}.

\section{Related Work}
\label{sec:mmab:related-work}

The MAB problems recently have attracted lots of attention in many application domains, such as opportunistic spectrum access~\cite{li2020multi}, and task offloading for vehicular edge computing~\cite{zhang2020adaptive}.
In the centralized setting, \cite{lai1985asymptotically} first provides an optimal
logarithmic lower bound on the regret of MAB algorithms,
then~\cite{anantharam1987asymptotically} extends this result to the setting
where several arms can be selected simultaneously at each time.

The distributed MMAB setting is first introduced
in~\cite{anandkumar2011distributed} and~\cite{liu2010distributed}. 
They design algorithms with fairness constraints and maintained an
order-optimal regret, which is further improved by~\cite{besson2018multi}
and~\cite{rosenski2016multi}. 
In these works, a player does not explicitly communicate with other players.
There has been a line of works that follow the same setting. 
In~\cite{gai2014distributed} and~\cite{kalathil2014decentralized}, the players
select different arms in the best set of arms according to their ranks. 
This approach is similar to ours and effective in avoiding collisions. 
\cite{bistritz2018distributed} proposes the first algorithm that enables
coordination among palyers and achieves a poly-logarithmic regret. 
\cite{shahrampour2017multi} considers an MMAB model where the expected reward
of each arm may vary from one palyer to another and a single group decision is
obtained through a voting process.
Our model is different as we consider each palyer makes a decision instead of a
group decision at each time.
\cite{boursier2019sic} and~\cite{wang_optimal_2020} use collisions as a means
of communication between palyers and achieved the best regret upper bound so far.
\cite{lugosi2022multiplayer} focus on a more challenging scenario
where the palyers do not get any information about a collision.

Recently, there has been a growing interest in distributed cooperative MMAB problems, where multiple players are able to collaboratively learn the arm reward distributions through limited communication.
\cite{tao2019collaborative} develops a collaborative learning algorithm
with limited communication steps based on the round elimination technique. 
\cite{landgren2018social} and~\cite{landgren2016distributed_2} propose Coop-UCB
and Coop-UCB2 algorithms, where players can collect information from their neighbors via a
running consensus procedure. 
A similar idea has also been suggested in the DD-UCB algorithm proposed
in~\cite{martinez2019decentralized}. 
However, DD-UCB requires extra knowledge of the spectral gap of the graph to
accelerate communication. 
\cite{zhu2021distributed} further considers a communication graph that only
allows uni-directional communication between neighboring palyers.
However, these works all considered MMAB models without collisions.
\cite{landgren2021distributed} designs the collaborative learning algorithms with collisions and without collisions and investigates the influence of the communication graph structure on performance. 
Our approach differs from previous studies in that we consider
fair arm (sensor) selection among players (servers) and we do not require prior knowledge of the
number of players in our model.


\section{Model Description}
\label{sec:syst-model-perf}

Consider an intelligent IoT system with a set of 
$\mathcal{M}=\{1,\ldots,M\}$ edge servers and a set of $\mathcal{N}=\{1,\dots,N\}$
$(N>M)$ sensors. 
At time $t \in \left\lbrace 1,\ldots,T\right\rbrace$, each server $k\in
\mathcal{M}$ select a sensor $i^{k}(t)\in \mathcal{N}$ for data collection.
Let $p^k_i(t)$ be the indicator that is equal to $1$
if server $k$ selects sensor $i$ at time $t$ and is equal to $0$ otherwise.
We denote random variable $a^{k}_{i}(t) \in[0,1]$ as the true data rate during the data transmission from sensor $i$ to server $k$. 
We assume that $a^{k}_{i}(t)$ evolves as an i.i.d. random process over
time with $\Ex[a^{k}_{i}(t)]=\mu_{i}$ for $k\in \mathcal{M}$ and $i\in \mathcal{N}$. 
In practice, $\mu_{i}$ can characterize the average data rate of sensor $i$ and is unknown to all servers.
Without loss of generality, we assume that $\mu_{1} \geq \mu_{2} \geq \ldots \geq \mu_{N}$.
Define the average data rate vector $\boldsymbol{ \mu } = \left[
\mu_{1},\ldots,\mu_{N} \right]$.
The servers can learn the vector $\boldsymbol{ \mu }$ cooperatively and share
estimations with their neighbors over an undirected and connected network
$G(\mathcal{M},\mathcal{E})$, where $\mathcal{M}$ represents the set of nodes
(servers) and $\mathcal{E}$ represents the set of edges (neighboring links).
There is an edge $kk'\in \mathcal{E}$ if server $k'$ is a neighbor of server $k$.
Each server can only communicate with its neighbors in $G$ at each time.
Collisions can occur when multiple servers select the same sensor at the same
time. 
Let $r_{i}^{k}(t)\equiv a_{i}^{k}(t)\eta^{k}(t)$ be the reward of server $k$
selecting sensor $i$, where $\eta^{k}(t)$ is a no-collision
indicator.
$\eta^{k}(t)=1$ if server $k$ does not collide with other servers at time $t$, and $0$ otherwise. 
Note that $\Ex[r^{k}_{i}(t)]=\mu_{i}\eta^{k}(t)$.
Thus, the colliding servers receive no rewards when a collision occurs.
We assume that the colliding servers can observe the selected sensors' data rates.
For example, if server $k_{1}$ and server $k_{2}$ select the same sensor at time
$t$, i.e., $i^{k_{1}}(t)=i^{k_{2}}(t)$, both servers receive zero reward, i.e.,
$r^{k_{1}}_{i}(t)=0$ and $r^{k_{2}}_{i}(t)=0$, but they can observe the their true data rate, $a^{k_{1}}_{i}(t)$ and $a^{k_{2}}_{i}(t)$.

Our main objective is to design a distributed sensor selection algorithm for
all servers to maximize their cumulative reward. 
We adopt the \emph{reward regret} from the MAB literature to measure the
performance of the algorithm.
The reward regret is defined as the cumulative difference between the expected
optimal reward and the expected reward of the algorithm. 
The expected optimal reward is $\sum_{k=1}^{M} \mu_k$, corresponding to the case
that $M$ servers select the $M$ sensors with the top $M$ average data rates
with no collisions. 
The expected reward of the algorithm is the reward received by the servers who
have not collided, i.e., $\sum_{k=1}^{M} \mu_{i^k(t)}\eta^{k}(t)$.
Then the \emph{reward regret} for the algorithm can be expressed as follows,
\begin{equation}
	\label{eq:mmab:dist-regret}
	\text{RR}_T=  T \sum_{k=1}^{M} \mu_k - \sum_{t=1}^{T} \sum_{k=1}^{M} \mu_{i^k(t)}\eta^{k}(t).
\end{equation}
The reward regret of all servers in \eqref{eq:mmab:dist-regret} can be decomposed into
the reward regrets of $M$ single servers, i.e., $\text{RR}_T=\sum_{k=1}^{M}\text{RR}^k_T = \sum_{k=1}^{M}(T \mu_k - \sum_{t=1}^{T} \mu_{i^k(t)} \eta^{k}(t))$.

Moreover, we also aim to ensure fairness among servers and guarantee all
servers receive the same amount of reward in expectation. 
To measure the fairness of an algorithm, we compute the absolute difference
between the average expected reward of all servers and the expected reward of
every server and define the \emph{fairness regret} as follows:
\begin{equation}
\label{eq:mmab:fairness-metric}
\text{FR}_T = \sum_{k=1}^{M}\Big|\sum_{t=1}^{T} \big[\Bar{\mu}(t)-\mu_{i^k(t)}\eta^{k}(t)\big]\Big|,
\end{equation}
where $\Bar{\mu}(t)=\frac{1}{M} \sum_{k=1}^{M}
\mu_{i^k(t)}\eta^{k}(t)$.
As proved in~\cite{anantharam1987asymptotically}, the instance-dependent regret
upper bound (depending on $\boldsymbol{\mu}$) for any bandit algorithm is
logarithmic, i.e., $O(\log T)$. 
Therefore, a good algorithm should have logarithmic instance-dependent reward
regret upper bound and fairness regret upper bound.

\section{Algorithm Design}
\label{sec-distributed-policy}

In this section, we outline our algorithm design and analyze its reward/fairness
regret.

\subsection{Initialization Scheme}
\label{sec:mmab:initialization}

In the beginning, the number of servers $M$ in the communication graph is unknown
to all servers. 
To estimate the value of $M$ for each server, we devise an initialization scheme
INIT based on~\cite{boursier2019sic} and assign a unique rank $h_0^k$ to server
$k$ for $k\in\mathcal{M}$. 

First of all, each server runs the Musical Chair
algorithm~\cite{rosenski2016multi}, and selects a sensor with the sensor index
being its rank. 
Next,
server $k$ can determine the number of servers $M$ and its rank $h_0^k\in
\mathcal{M}$ by observing the number of collisions. 
Due to the space limit, we put detailed descriptions of the initialization
scheme in Appendix~A in our technical report~\cite{TechnicalReport}, where we
also prove the following theorem.
\begin{myTheo} 
\label{the:initialization-scheme-time-slots}
\label{lem} With probability at least $1-\delta_0$, by running the
initialization scheme $\text{INIT}(N, \delta_0)$, all servers can learn $M$ and
get a distinct rank from $1$ to $M$ using $N \ln(e^2N / \delta_0)$ time slots.
\end{myTheo}

From Theorem~\ref{lem}, it follows that the number of time slots required to
complete the initialization scheme is fixed with a given $\delta_{0}$. 
Thus, the initialization scheme only incurs a negligible constant amount of
reward/fairness regret. 

\subsection{Running Consensus for Data Rate Estimation}
\label{sec:mmab:running_consensus}

When each server gets the number of servers $M$ after the initialization
scheme, we can establish a communication protocol between servers for efficient
information dissemination over the network $G$. 
Our goal is to help each server obtain global observations to cooperatively
estimate the expected data rates of all sensors.

Running consensus~\cite{braca2008enforcing} is an effective method to keep
running approximations of the expected values of random variables.
In the cooperative estimation of the expected data rate of sensor $i$, we denote
$\hat{g}^k_{i}(t)$ and $\hat{n}^k_{i}(t)$ as server $k$'s estimate of the average
data rate from sensor $i$ up to time $t$, and the number of times that
sensor $i$ has been selected up to time $t$, respectively. 
The estimates $\hat{g}^k_{i}(t)$ and $\hat{n}^k_{i}(t)$ can be updated with
a running consensus procedure as follows:
\begin{equation} 
\hat{\bm{g}}_{i}(t)=\bm{S}\left(\hat{\bm{g}}_{i}(t-1)+\bm{a}_{i}(t)\right),
\label{eq:mmab:reward-per-user}
\end{equation}
\begin{equation}
\hat{\bm{n}}_{i}(t)=\bm{S}\left(\hat{\bm{n}}_{i}(t-1)+\bm{p}_{i}(t)\right), 
\label{eq:mmab:samples-per-user}
\end{equation}
where $\hat{\bm{g}}_{i}(t)$, $\hat{\bm{n}}_{i}(t)$, $\bm{a}_{i}(t)$,
and $\boldsymbol{p}_{i}(t)\in \mathbb{R}^{M \times 1} $ are vectors of
$\hat{g}_{i}^{k}(t)$, $\hat{n}_{i}^{k}(t)$, $a_{i}^{k}(t)$, and $p_{i}^{k}(t)$,
and $\bm{S}\in \mathbb{R}^{M \times M} $ is a gossip matrix representing the
structure of the network and $\bm{S}_{kk'}=0$ if there is no neighboring link between server $k$
and server $k'$.
We construct $\bm{S}$ in a way such that each server's observations can converge to
the average of all servers' observations across the entire network $G$ over time.
In other words, the $t$-th power of the gossip matrix $\bm{S}$ can converge to a matrix whose
elements are all $\frac{1}{M}$, i.e., $\lim \limits_{t \to \infty} \bm{S}^t =
\frac{\bm{1}\bm{1}^T}{M}$ 
where $\bm{1} \in \mathbb{R}^{M \times 1}$ is the one vector.
In the convergence state, the matrix $\bm{S}$ is a doubly
stochastic matrix
where the sum of each row and the sum of each column is $1$,
and all eigenvalues of $\bm{S}$ are real and less than $1$.
See \cite{xiao2004fast} for the proof and \cite{xiao2004fast,duchi2011dual} for discussions on how to choose $\bm{S}$.

Thus, revisiting \eqref{eq:mmab:reward-per-user} and
\eqref{eq:mmab:samples-per-user}, we can describe the running consensus
procedure in detail as follows. 
Server $k$ collects the values of $a^k_i(t)$ and $p^k_i(t)$ for sensor $i$ at
time $t$ and adds them to the current estimates $\hat{g}_{i}^{k}(t-1)$ and
$\hat{n}_{i}^{k}(t-1)$ respectively, then uses the gossip matrix $\bm{S}$ to get
the updated estimates: $\hat{g}_{i}^k(t)$ and $\hat{n}_{i}^k(t)$. 
When $t \rightarrow \infty$, we have $\hat{g}_{i}^k(t) \rightarrow
\overline{g}_i(t) =\frac{1}{M} \sum_{k=1}^{M}\sum_{\tau =1}^{t} a^k_i( \tau )$
and $\hat{n}_{i}^k(t) \rightarrow
\overline{n}_i(t)=\frac{1}{M}\sum_{k=1}^{M}\sum_{\tau =1}^{t}p^k_i(\tau)$
according to the convergence property of $\bm{S}$. 
Then server $k$ can estimate the expected data rate of sensor $i$ at time $t$ with 
$\hat{\mu} _{i}^{k}(t) = \frac{\hat{g}_{i}^{k}(t)}{\hat{n}_{i}^{k}(t)}$.
To describe the connectivity of the network $G$, we introduce a graph structure index $\epsilon_g$,
$\epsilon_g=\sqrt{M} \sum_{x=2}^{M} \frac{\left|\lambda_{x}\right|}{1-\left|\lambda_{x}\right|}$,
where $\lambda_{x}$ is the $x$-th largest eigenvalue of gossip matrix
$\bm{S}$.
Note that $\epsilon_g$ decreases
as $| \lambda_x |$ decreases for any $x \geq 2$.

Via the running consensus procedure, each server can share the information,
namely, $\bm{a}_{i}(t)$ and $\bm{p}_{i}(t)$, with its neighbors in $G$ at
time $t$. 
Using such information, every server can estimate the average number of
selected times $\hat{n}_{i}^{k}(t)$ and the average data rate $\hat{\mu}
_{i}^{k}(t)$ for $k\in\mathcal{M}$.
We have the following results on the accuracy of $\hat{n}_{i}^{k}(t)$ and
$\hat{\mu} _{i}^{k}(t)$.
\begin{lemma}
\label{pro:mmab:performance-running-consensus} With the running consensus
procedure,
for $k\in\mathcal{M}$, the estimate $\hat{n}_{i}^{k}(t)$ satisfies
\begin{equation}
\label{eq:mmab:selection-estimate-interval}
\overline{n}_{i}(t)-\epsilon_g \leq \hat{n}_{i}^{k}(t) \leq \overline{n}_{i}(t)+\epsilon_g \text{,}
\end{equation}
and the estimate $\hat{\mu} _{i}^{k}(t)$ is unbiased.
\end{lemma}
We refer interested readers to Appendix~B in our technical
report~\cite{TechnicalReport} for the proof. 
Note that a smaller $\epsilon_g$ indicates a higher level of connectivity in the
graph, allowing each server to collect more information about the rewards of the
sensors and thereby providing more accurate estimates of $\hat{n}_{i}^{k}(t)$
and $\hat{\mu} _{i}^{k}(t)$.

\subsection{\dcul{} Algorithm}
\label{ssec:mmab:dcul-algorithm}
\addtolength{\topmargin}{0.01in}
Given the sensor's data rate estimates by the running consensus procedure, we can
then finally develop our distributed cooperative bandit algorithm \dcul{}.

To avoid collisions, the optimal sensor selection algorithm is to have each of
the $M$ servers selecting a different sensor from the $M$ best sensors with
expected data rates $\left\lbrace \mu_{1},\ldots,\mu_{M} \right\rbrace $. 
Therefore, we can force a server to select a different sensor from the set of $M$
best sensors according to the sensor reward estimates and its rank at
each time. 
For instance, server $k$ selects the $h_0^k$-th best sensor with the
$h_0^k$-th largest reward estimate, but it results in an unfair sensor selection
where the servers with smaller ranks always have higher priorities in selecting a
sensor with larger reward estimates. 
To satisfy the fairness requirement, we treat servers equally by changing the rank
$h^k(t)$ for each server $k$ cyclically in the set $\{ 1,..., M\}$, i.e.,
$h^k(t)=((h_0^k+t) \mod M) +1$. 
In this way, we can achieve a fair sharing of the $M$ best sensors among $M$
servers.

At this point, server $k$ selects the sensor with the $h^k(t)$-th largest reward
estimate. 
Therefore, the traditional UCB (upper confidence bound) algorithm where the server just
selects the sensor with the highest UCB estimate is inapplicable. 
To this end, we consider estimating the expected reward of sensors by using the
UCB and LCB (lower confidence bound) of sensors
simultaneously. 
The UCB estimate and LCB estimate of the reward of sensor $i$ for server $k$ are
computed as follows:
\begin{equation}
    U_{i}^{k}(t)=\hat{\mu} _{i}^{k}(t-1)+C_{i}^k(t-1),
    \label{eq:mmab:ucb}
\end{equation}
\begin{equation}
    L_{i}^{k}(t)=\hat{\mu} _{i}^{k}(t-1)-C_{i}^k(t-1),
    \label{eq:mmab:lcb}
\end{equation}
where $\hat{\mu} _{i}^{k}(t) = \frac{\hat{g}_{i}^{k}(t)}{\hat{n}_{i}^{k}(t)}$ is
the estimate of the expected data rate of sensor $i$ using the running consensus method and $C_{i}^k(t)=\sqrt{\frac{2 \ln
Mt}{M\hat{n}^k_i(t)}}$ is the radius of a confidence interval around $\hat{\mu}
_{i}^{k}(t)$ at time $t$. Note that $\mu_i \in [L_{i}^{k}(t),
U_{i}^{k}(t)]$ holds with high probability according to the concentration
properties of random variables. $\hat{\mu} _{i}^{k}(t)$ and
$C_{i}^k(t)$ are updated with the running consensus procedure.

\setlength{\textfloatsep}{3pt}
\begin{algorithm}[!t]
\renewcommand{\algorithmicrequire}{\textbf{Input:}}
\renewcommand{\algorithmicensure}{\textbf{Initialization:}}
\caption{\dcul{} at Server $k$}
\label{alg:mmab:Dco-ULCB}
\begin{algorithmic}[1]
\REQUIRE $N$ (number of sensors), $T$ (time horizon).
\ENSURE  $M, h_0^k \leftarrow \text{INIT}(N, \frac{1}{NT})$.
\STATE \textbf{set} $\hat{g}_{i}^{k}(0) \leftarrow 0,\hat{n}_{i}^{k}(0) \leftarrow 0$ for $i \in \mathcal{N}$; 
\FOR{$t=1$ to $T$}
	\IF{$t \leq N$} 
	   \STATE $i^{k}(t) \leftarrow ((h_0^k+t) \mod N) +1$;
	   \STATE Observe $p^k_i(t)$, $a^k_i(t)$, $\eta^k(t)$  and receive reward $r^{k}_i(t)$;            
	\ELSE
	   \STATE \COMMENT{Change the rank cyclically}\\ $h^k(t) \leftarrow ((h_0^k+t) \mod M)+1$; 
	   \FOR{sensor $i \in \mathcal{N}$}
	   \STATE
       Compute UCB $U_{i}^{k}(t)$ using \eqref{eq:mmab:ucb};
	   \ENDFOR
       \\\COMMENT{Descending sort by UCB}
	   \STATE $\mathcal{R}_{h^k(t)} \leftarrow \{ I^{(1)},..., I^{(h^k(t))} \} $;
	   \FOR{sensor $i \in \mathcal{R}_{h^k(t)}$}
	   \STATE
       Compute LCB $L_{i}^{k}(t)$ using \eqref{eq:mmab:lcb};
	   \ENDFOR
	   \STATE $i^k(t) \leftarrow \mathop{\arg\min}_{i \in \mathcal{R}_{h^k(t)}}  L_{i}^{k}(t) $; 
	   \STATE Observe $p^k_i(t)$, $a^k_i(t)$, $\eta^k(t)$  and receive reward $r^{k}_i(t)$;            
    \ENDIF 
    \STATE Update $\hat{g}_{i}^{k}(t)$, $ \hat{n}_{i}^{k}(t)$ for $i \in \mathcal{N}$ using \eqref{eq:mmab:reward-per-user} and \eqref{eq:mmab:samples-per-user};     
	\ENDFOR
\end{algorithmic}
\end{algorithm}

Our \dcul{} algorithm incorporating the running consensus procedure is described
in Algorithm~\ref{alg:mmab:Dco-ULCB}. 
Each server first estimates the number of servers $M$ and gets a unique rank $h_0^k$ via
the initialization scheme. 
From line $3$ to line $5$, server $k$ selects each sensor once
to obtain the initial values of $p^k_i(t)$ and $a^{k}_i(t)$. The updated rank 
$h^k(t)$ is calculated at line $7$. 
Then server $k$ selects the sensor with the $h^k(t)$-th largest reward estimate. 
We first use the UCB value as an estimate of the reward of a sensor. 
Server $k$ constructs the set $\mathcal{R}_{h^k(t)}= \{ I^{(1)},..., I^{(h^k(t))}
\}$ containing the indices of the sensors with the top $h^k(t)$ largest UCB
values at line $11$. 
Then we consider the LCB value as another estimate of the reward of a sensor.
Therefore, the $h^k(t)$-th best sensor can be obtained by selecting the sensor
with the smallest LCB value in $\mathcal{R}_{h^k(t)}$. 
The algorithm updates $\hat{g}_{i}^{k}(t)$ and $ \hat{n}_{i}^{k}(t)$ by the
running consensus procedure using \eqref{eq:mmab:reward-per-user} and
\eqref{eq:mmab:samples-per-user} at line $18$. 

In contrast to~\cite{landgren2021distributed}, we enforce a fair sensor
selection among servers and consider a more challenging setting with no prior
knowledge of the number of servers.
Moreover, our algorithm generalizes the DLF algorithm~\cite{gai2014distributed} that
strictly assumes no communication among servers to the case of communicating servers. 
In addition, our algorithm also works in the fully distributed MMAB setting where
there is no communication link between any two servers. 
In this case, we have $\bm{S}_{kk'}=0, \forall k \neq k'$ and each server
$k$ only updates $\hat{g}_{i}^{k}(t)$ and $ \hat{n}_{i}^{k}(t)$ for $i \in
\mathcal{N}$ based on its local observations.

\subsection{Regret Analysis}
\label{ssec:mmab:regret-analysis}

In this section, we analyze the reward/fairness regret upper bound of \dcul{} using the results
in Lemma~\ref{pro:mmab:performance-running-consensus}.
To proceed, we first denote $\tilde{m}^k_i(t)$ as the number of times that
sensor $i$ has been incorrectly selected by server $k$ up to time $t$. 
The incorrect sensor selections are the primary cause of regret. 
We can bound the number of incorrect sensor selections in the following
theorem.
\begin{myTheo}
\label{the:mmab:subopt-upper-bound}
By running the \dcul{} algorithm, the expected
number of times that all the servers incorrectly select sensor $i$ until time $T$
satisfies:
\begin{equation}
\label{eq:mmab:subopt-upper-bound}
\sum_{k=1}^{M}\Ex[\tilde{m}^k_i(T)] \leq \dfrac{8 \ln MT}{\Delta^2_{ \min }}+M \epsilon_g+\frac{2 \pi^{2}}{3M^3}+1 ,
\end{equation}
where $\Delta_{ \min }= \min \{\left| \mu_{i} - \mu_{j} \right|  \mid i,j \in \mathcal{N},\mu_{i} \neq \mu_{j} \}$. 
\end{myTheo}
We refer the interested readers to Appendix C in~\cite{TechnicalReport}
for the full proof. 
There are two cases that increase the reward regret at time $t$ for
server $k$.
One is that server $k$ selects sensor $i \neq h^k(t) $ and the other is that
server $s \neq k$ selects sensor $h^k(t)$. 
Collisions may happen in both situations and increase the reward regret
by at most $1$. 
Hence, the reward regret of server $k$ until time $T$ can be divided into two parts:
\begin{equation}
\label{eq:mmab:Dco-ULCB-regret-bound}
    \text{RR}_{T}^{k} \leq \sum_{i=1}^{N}  \Ex[\tilde{m}_i^k(T)] + \sum \limits_{s \neq k}  \sum_{i=1}^{M} \Ex[\tilde{m}_{i}^s(T)]\text{.}
\end{equation}
Let $Z(\bm{\mu},T,\epsilon_g)\equiv\dfrac{8 \ln MT}{\Delta^2_{ \min }}+M \epsilon_g+\frac{2 \pi^{2}}{3M^3}+1$.
Combining with \eqref{eq:mmab:subopt-upper-bound} and
\eqref{eq:mmab:Dco-ULCB-regret-bound},
the upper bound of the reward regret of \dcul{} satisfies:
\begin{align}
&\text{RR}_T {=} \sum \limits_{k=1}^M \text{RR}_{T}^{k}\nonumber 
\leq \sum_{i=1}^{N}  \sum_{k=1}^{M}\Ex[\tilde{m}_i^k(T)] {+} \sum_{k=1}^{M} \sum_{i=1}^{M} \sum \limits_{s \neq k}\Ex[\tilde{m}_{i}^s(T)]\nonumber \\
&\leq \sum_{i=1}^{N} Z(\bm{\mu},T,\epsilon_g) + \sum_{k=1}^{M} \sum_{i=1}^{M} Z(\bm{\mu},T,\epsilon_g) \nonumber \\
&\leq (N+M^2)Z(\bm{\mu},T,\epsilon_g).
\label{eq:mmab:Dco-ULCB-system-regret-bound}
\end{align}
\eqref{eq:mmab:Dco-ULCB-system-regret-bound} shows that the reward regret
of \dcul{} is logarithmic in time $T$ and depends
on the graph index $\epsilon_g$, which achieves order-optimal regret.  
Note that \dcul{} ensures that servers receive the same expected reward if they
correctly select all sensors. 
Thus, the upper bound of fairness regret of the DC-ULCB algorithm satisfies:
\begin{equation}
\label{eq:mmab:fairness-metric-bound}
\text{FR}_T   \leq   \sum_{i=1}^{N}\sum_{k=1}^{M}\Ex[\tilde{m}^k_i(T)]   \leq NZ(\bm{\mu},T,\epsilon_g),
\end{equation}
which is similar to (\ref{eq:mmab:Dco-ULCB-system-regret-bound}) and also
logarithmic in time $T$.

Thus, the instance-dependent reward regret upper bound in
\eqref{eq:mmab:Dco-ULCB-system-regret-bound} and the instance-dependent fairness
regret upper bound in \eqref{eq:mmab:fairness-metric-bound} are both logarithmic
in $T$. At the end of each round, each server needs to send information about all sensors to its neighbors,
incurring a communication cost $O(N)$.
In particular, the communication cost can be reduced at the expense of greater regret,
capturing the communication-regret tradeoff.

\section{Numerical Experiments}
\label{sec-numerical-experiments}

In this section, we conduct simulations to show the effectiveness of our \dcul{}
algorithm. 
We consider $N=40$ sensors with the expected data rates
$\boldsymbol{\mu}=\{\frac{i}{N+1}| i = 1, ..., N\}$. 
Each sensor $i$ transmits data with a random data rate sampled from a Beta distribution $\text{Beta}(\alpha_i,\beta_i)$ with mean $\mu_i$ where $\alpha_i=20, \beta_i=20\frac{1-\mu_i}{\mu_i}$.
We use the ER (Erdős–Rényi) model~\cite{ermodel} to generate a connected network with $M=10$
servers and connection probability $q=0.5$ between any two servers using lots of trials. 
Note that the generated network is kept fixed throughout our experiments unless
otherwise stated and all results are averaged over $100$ runs.

\begin{figure*}[!t]
\centering
\label{fig:mmab:Dco-ULCB}
\subfigure[Reward regret]{
	       \centering
	       \includegraphics[width=0.231\linewidth]{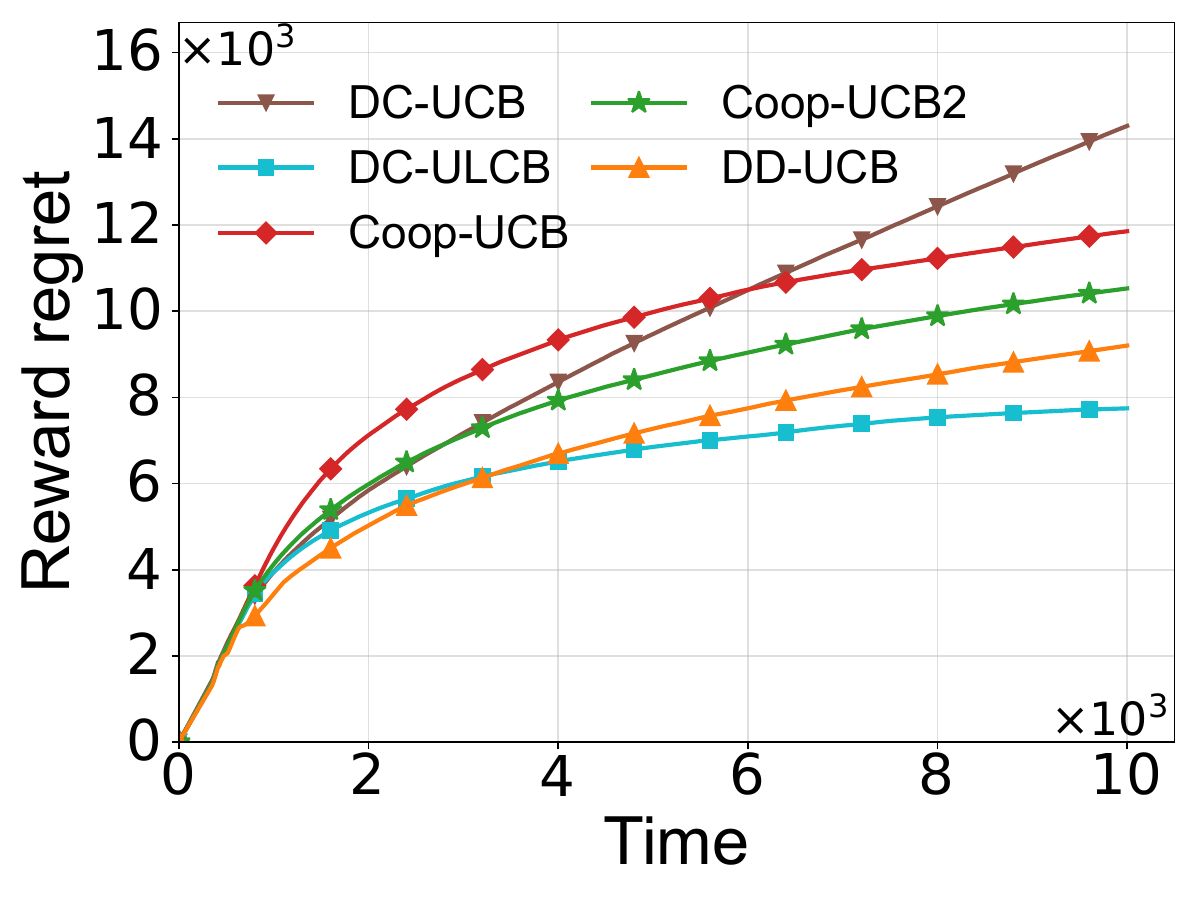}
              \label{fig:reward_regret_comparison_algorithm}              
    }
\subfigure[Fairness regret]{
	       \centering
	       \includegraphics[width=0.231\linewidth]{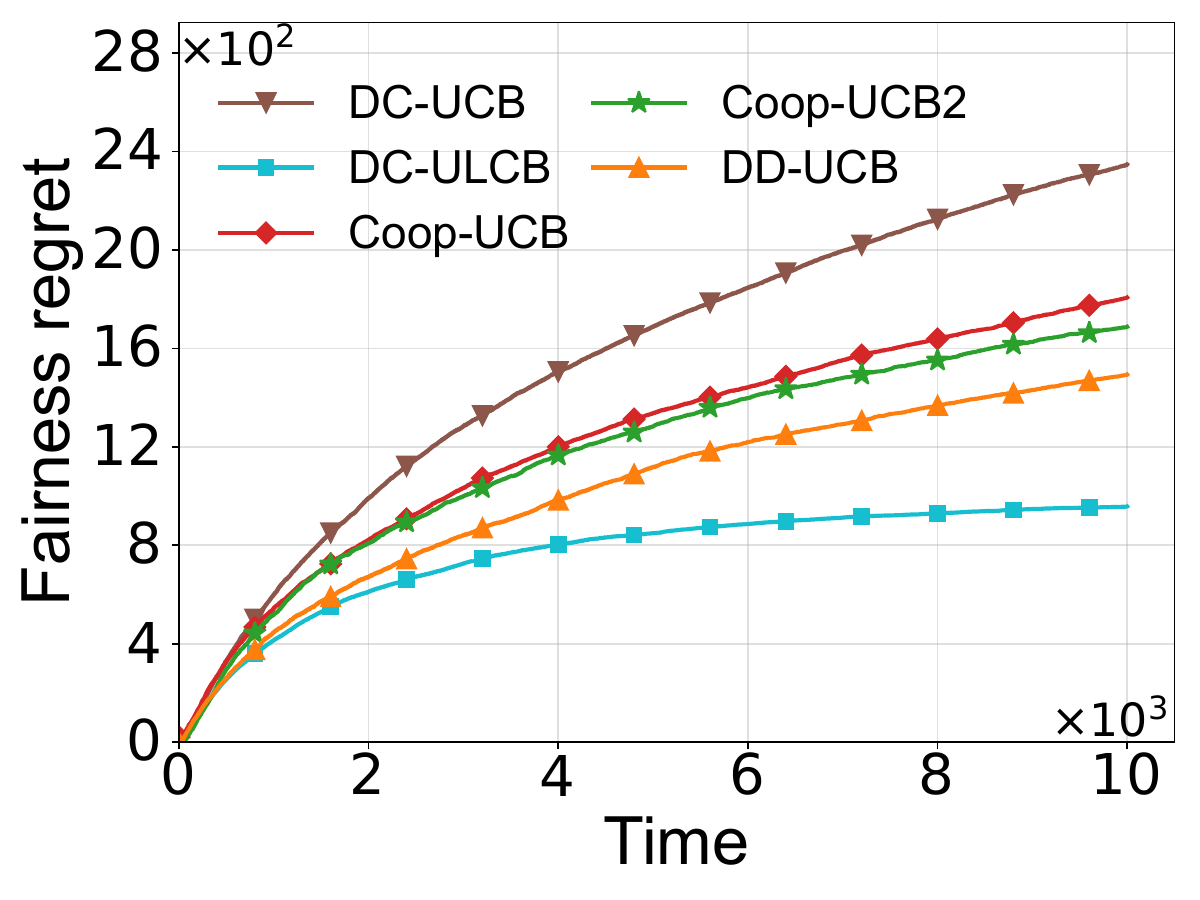}
              \label{fig:fairness_regret_comparison_algorithm}
    }
    \subfigure[Reward regret with different $q$]{
	    \centering
            \includegraphics[width=0.231\textwidth]{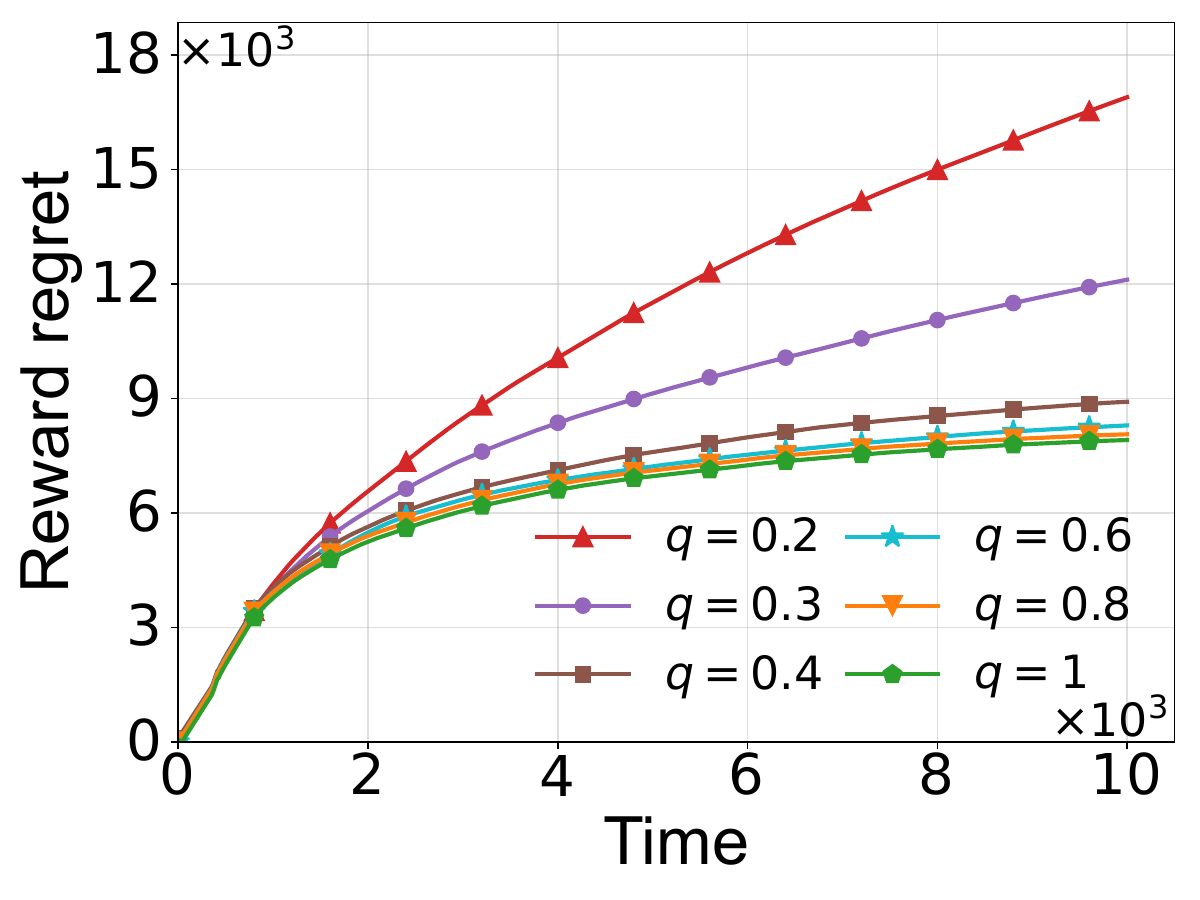}
            \label{fig:mmab:reward-regret-connection-pro}
    }
    \subfigure[Fairness regret with different $q$]{
	    \centering
            \includegraphics[width=0.231\textwidth]{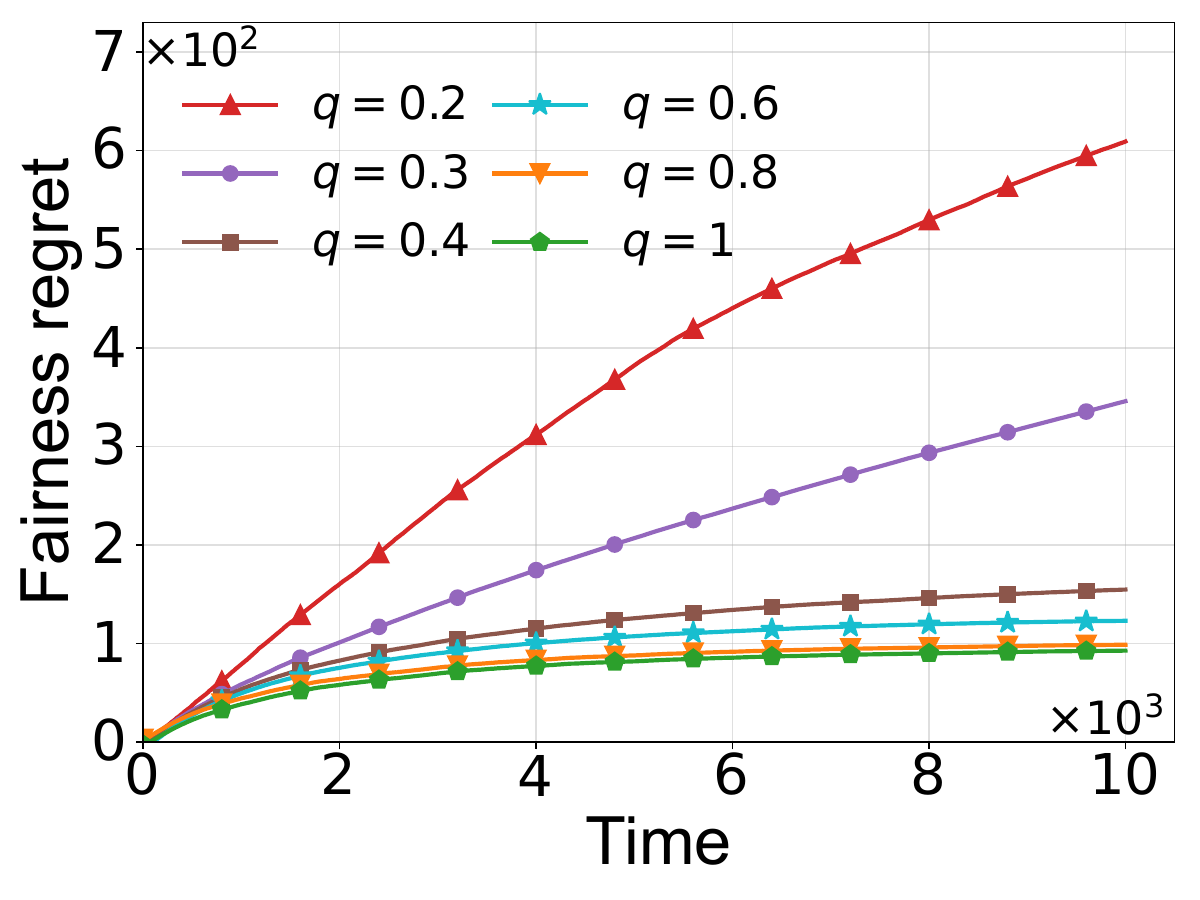}
            \label{fig:mmab:fairness-regret-connection-pro}
    }
\vspace*{-8pt}

    \subfigure[$\epsilon_g$ with different $q$]{
	    \centering
            \includegraphics[width=0.231\linewidth]{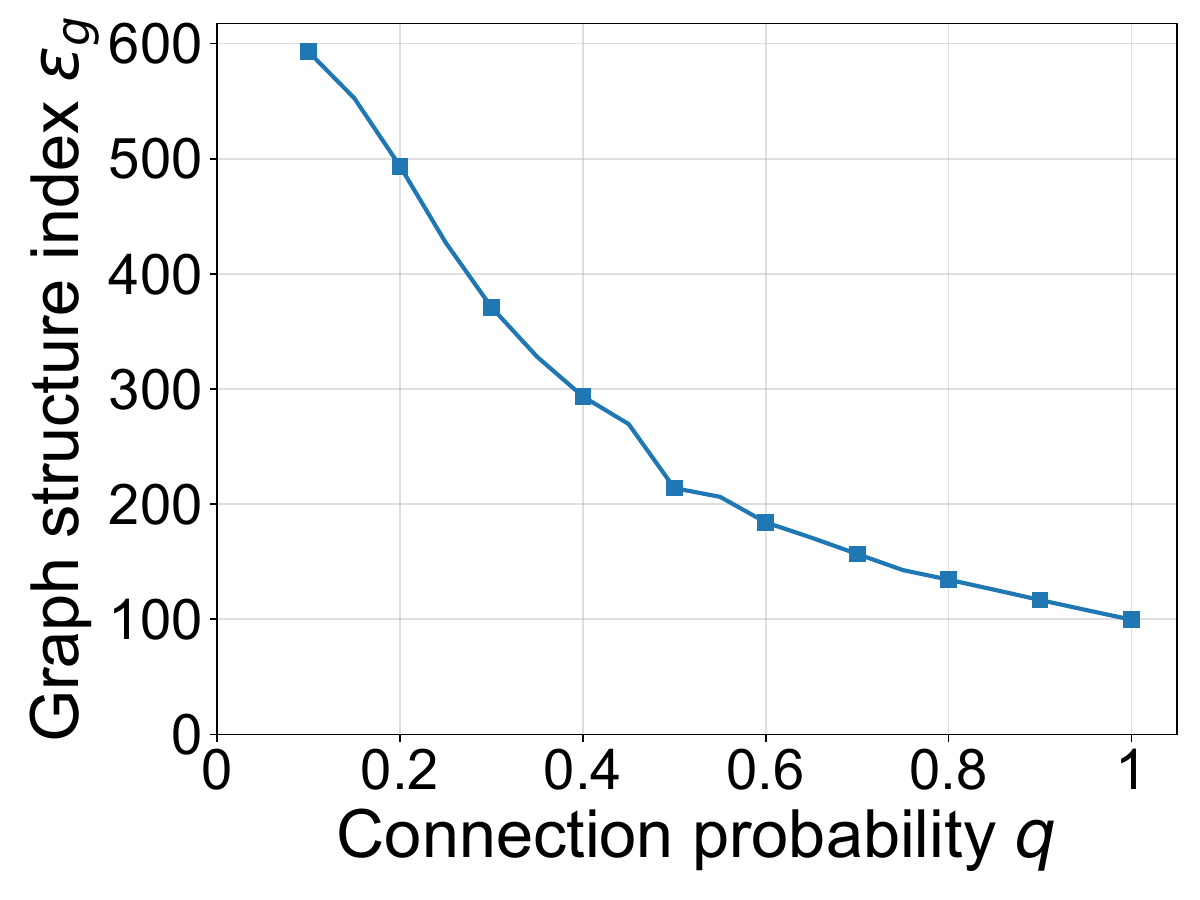}
            \label{fig:mmab:epsilon_g_revolution}
    }
    \subfigure[Average reward w/wo fairness]{
	    \centering             
            \includegraphics[width=0.231\textwidth]{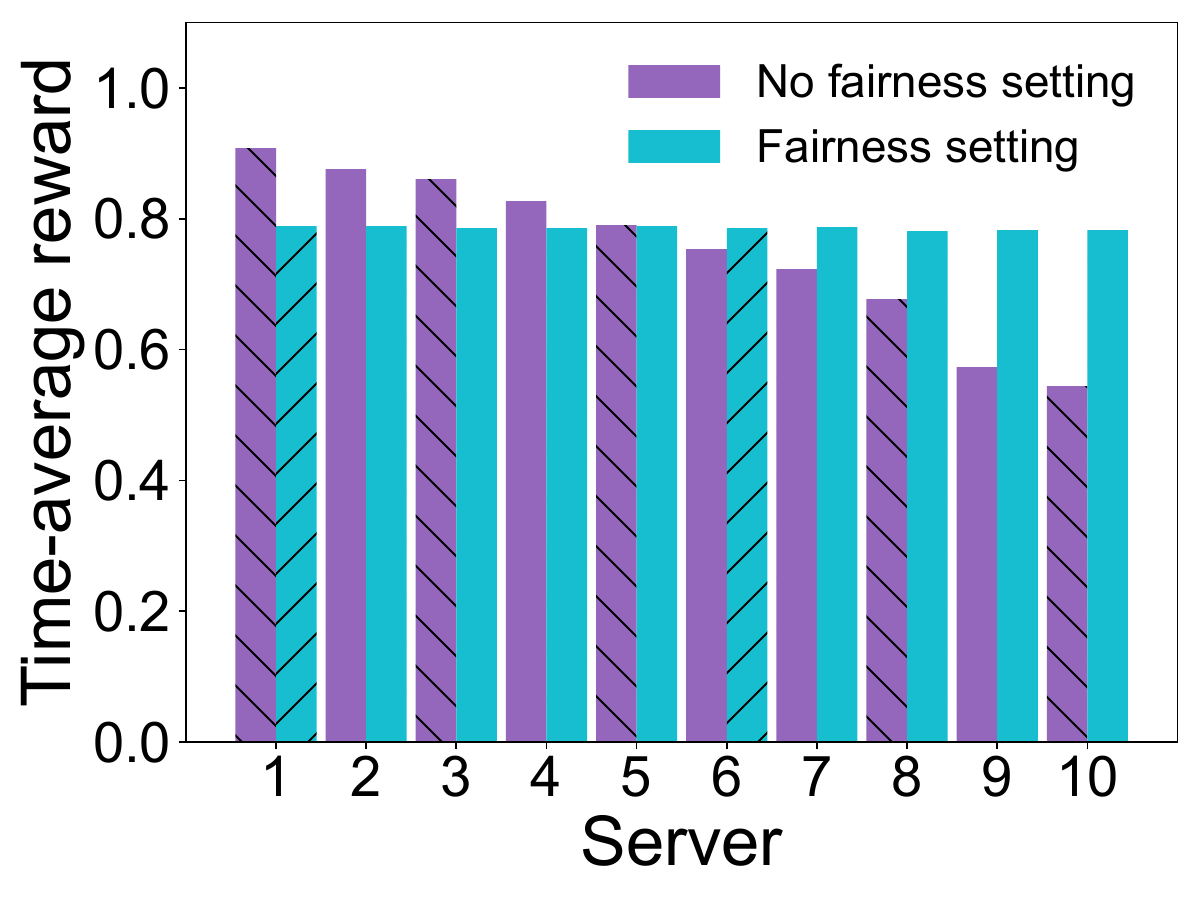}
            \label{fig:mmab:Average-reward-with-without-fairness}
    }
    \subfigure[Reward regret w/wo fairness]{
	    \centering
            \includegraphics[width=0.231\linewidth]{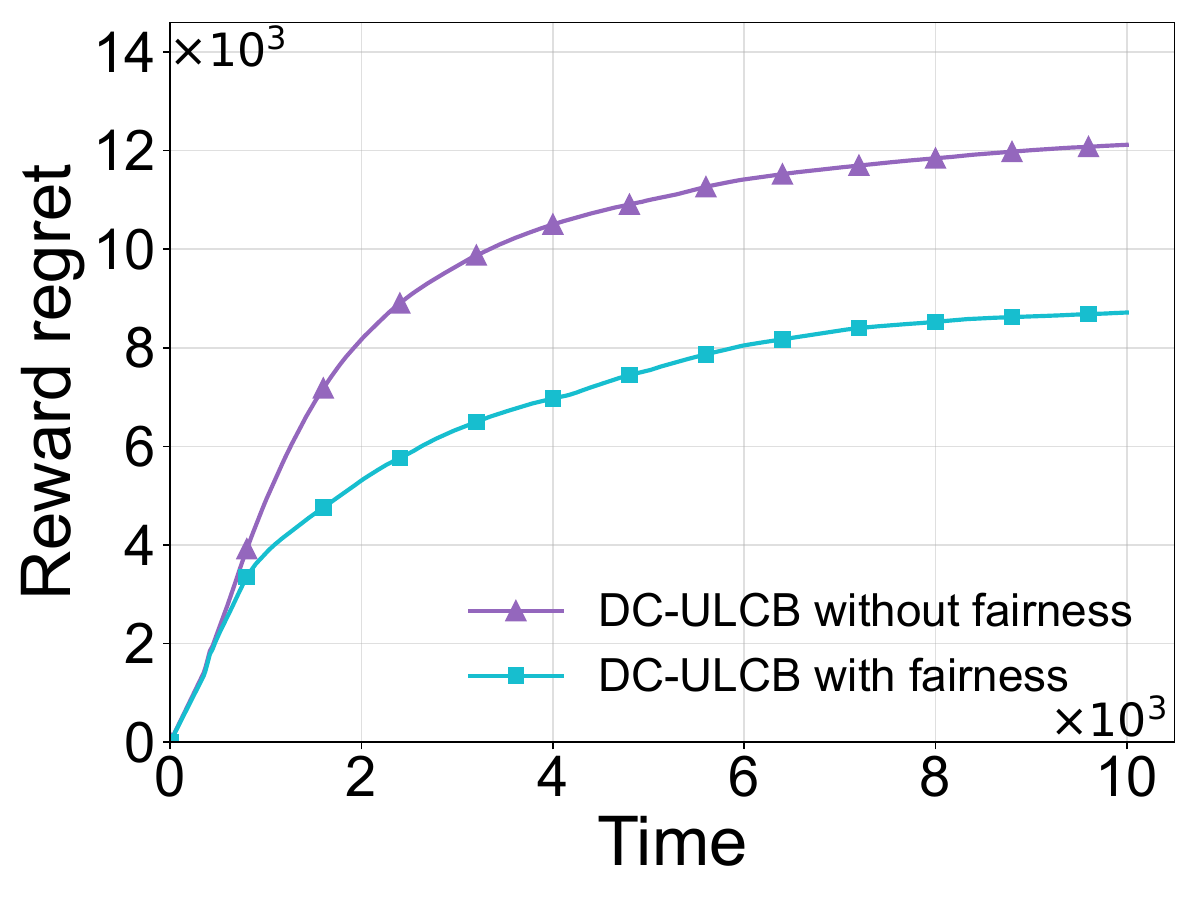}
            \label{fig:mmab:DC-ULCB-with-without-fairness-reward-regret}
    }
    \subfigure[Cum. collisions w/wo fairness]{
	       \centering
	       \includegraphics[width=0.231\linewidth]{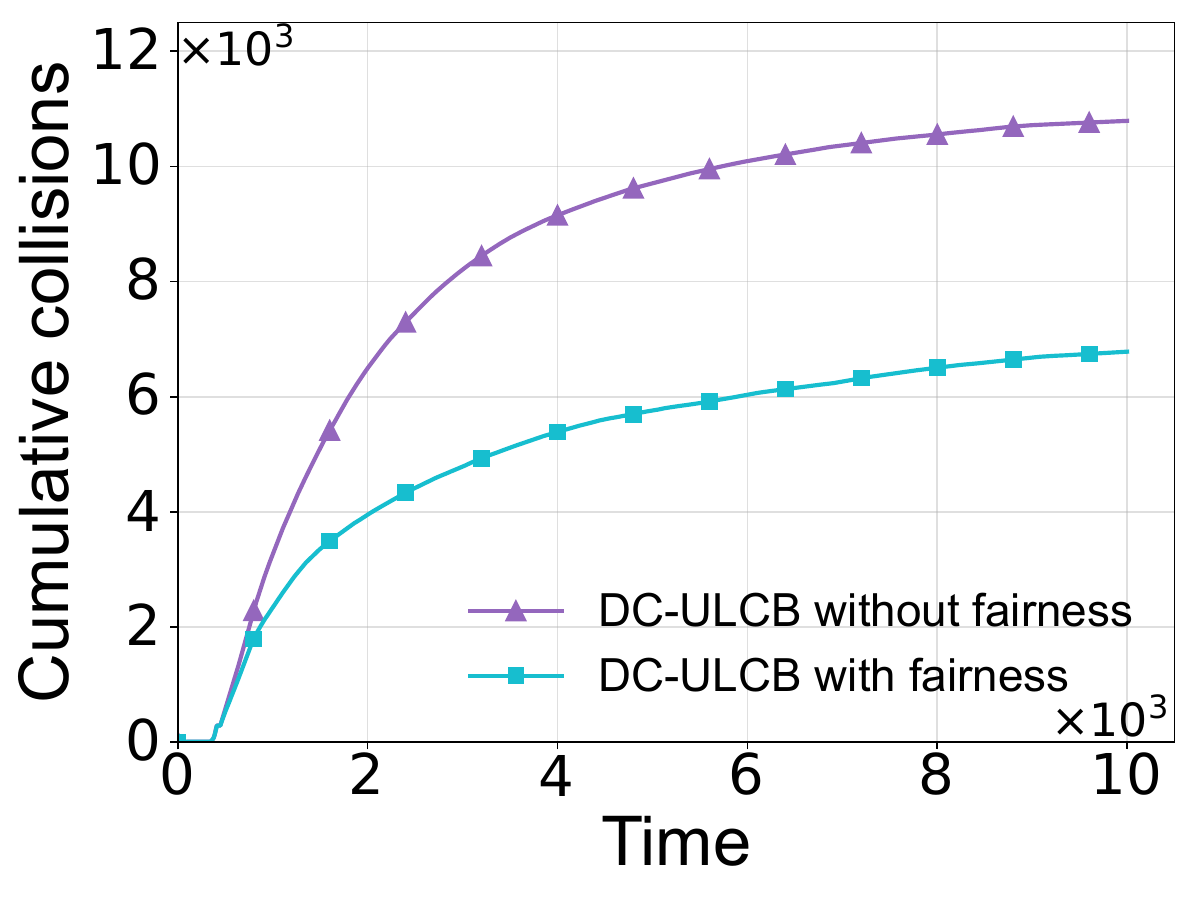}
	       \label{fig:mmab:DC-ULCB-with-without-fairness-cumulative-collisions}
    }
\caption{Experiment results on \dcul{}}
\vspace*{-10pt}
\end{figure*}

Fig.~\ref{fig:reward_regret_comparison_algorithm} and
Fig.~\ref{fig:fairness_regret_comparison_algorithm} show the reward regrets and
fairness regrets of different algorithms, respectively. 
In the naive DC-UCB (Distributed Cooperative UCB) algorithm, servers select the
sensors with the UCB only (without using the LCB) at each time $t$. 
As shown in Fig.~\ref{fig:reward_regret_comparison_algorithm} and
Fig.~\ref{fig:fairness_regret_comparison_algorithm}, \dcul{} has smaller
reward/fairness regret than DC-UCB. 
This is reasonable as \dcul{} uses the information in both the UCB and the LCB for
sensor selection to improve performance. 
We also observe that the reward/fairness regret of \dcul{} increases sublinearly in $T$
which is consistent with the $O (\log T)$ bound we derived
in~\eqref{eq:mmab:Dco-ULCB-system-regret-bound} and~\eqref{eq:mmab:fairness-metric-bound}. 
Moreover, we compare \dcul{} with the state-of-the-art algorithms, namely, Coop-UCB~\cite{landgren2018social},
Coop-UCB2~\cite{landgren2016distributed_2} and
DD-UCB~\cite{martinez2019decentralized} which are designed to accommodate the
collisions. 
Fig.~\ref{fig:reward_regret_comparison_algorithm} and Fig.~\ref{fig:fairness_regret_comparison_algorithm} show that \dcul{} significantly
outperforms all competitors in both reward regret and fairness regret.

Next, we study the influence of the connection probability $q$ of the ER graphs
on the performance of \dcul{} algorithm. 
We vary $q$ from $q=0.2$ (nearly fully distributed) to $q=1$ (centralized). 
For each $q$, we generated $20$ connected ER graphs. 
Fig.~\ref{fig:mmab:reward-regret-connection-pro} and Fig.~\ref{fig:mmab:fairness-regret-connection-pro} show that the reward/fairness regret
of \dcul{} decreases as $q$ increases.
This is consistent with our theoretical result in~\eqref{eq:mmab:Dco-ULCB-system-regret-bound} and~\eqref{eq:mmab:fairness-metric-bound}, where a higher $q$ results in smaller
$\epsilon_{g}$ (shown in Fig.~\ref{fig:mmab:epsilon_g_revolution}) and therefore
a lower regret.
Further observation shows that the improvement of reward/fairness regret is marginal when
$q$ is closer to $1$. 
Therefore, for real-world intelligent IoT systems, \dcul{} works
efficiently even if the network does not have a high connectivity.

Finally, we evaluate the performance of DC-ULCB with and without fairness
considerations. 
Note that if server $k$ selects the $h_0^k$-th best sensor in \dcul{}, \dcul{} would
lose fairness in sensor selection. 
Fig.~\ref{fig:mmab:Average-reward-with-without-fairness} shows the average
reward per time slot for each server without and with fairness over $T=10^4$, respectively.
This shows that each server does achieve the same time average reward under the
fairness setting in \dcul{}. 
Fig.~\ref{fig:mmab:DC-ULCB-with-without-fairness-reward-regret} further compares
the reward regrets of \dcul{} with and without fairness. 
The simulation result demonstrates that \dcul{} has a smaller reward regret
under the fairness setting. 
The reason is that the selected sensor of each server is cyclically changing in
the $M$ best sensors under the fairness setting, which can accelerate the
exploration process. 
Thus, the servers can learn information about each sensor faster compared to the
simple case where each server sticks to selecting only one sensor. 
As a result, the servers can estimate the sensor rewards more accurately and
avoid collisions earlier thereby resulting in a smaller reward regret. 
To verify, we show the cumulative number of collisions of \dcul{} without
fairness and with fairness in
Fig.~\ref{fig:mmab:DC-ULCB-with-without-fairness-cumulative-collisions}. 
As expected,
Fig.~\ref{fig:mmab:DC-ULCB-with-without-fairness-cumulative-collisions} shows
that \dcul{} with fairness has fewer collisions.

\section{Conclusion}
\label{sec-conclusion}

In our work, we propose a multiplayer multi-armed bandit model that considers both server
communications and fairness requirements for intelligent IoT systems.
We design a distributed cooperative bandit algorithm, \dcul{}, and help multiple
servers to select multiple sensors fairly and effectively.
With an initialization scheme, our \dcul{} algorithm that incorporates the
running consensus procedure can select proper sensors fairly and maximize the
total reward for the servers.
We further prove that \dcul{} can achieve order-optimal logarithmic
reward/fairness regret upper bounds. 
Our extensive simulation results also demonstrate the effectiveness of \dcul{}
compared to other algorithms.


\bibliographystyle{IEEEtran}
\bibliography{IEEEexample}
\clearpage
\appendices
\section{Description of the Initialization Scheme}
\label{sec:mmab:initialization-detail}
The initialization phase consists of two procedures. Let $T_0=N \ln N/ \delta_0$. The first procedure relies on the musical chair algorithm (Algorithm~\ref{alg:mmab:musical-chair}) introduced in \cite{rosenski2016multi}, where each server tries to find a distinct sensor that does not collide with other servers and gets an external rank during the first $T_0$ time slots. The probability of a server selecting a sensor that does not collide with other servers is at least $1/N$, since there exists at least one sensor that is not selected at each time ($N > M$ ). Therefore, the probability that a server does not succeed in getting an external rank during the $T_0$ time slots is upper bounded by 
\begin{equation}
\label{eq:mmab:failure-pro}
(1-1 / N)^{T_{0}}<\exp \left(-T_{0} / N\right)=\delta_{0} / N \leq \delta_{0} / M \text{,}
\end{equation}
Hence, the union bound over the $M$ servers proves that all servers successfully get an external rank with probability at least $1-\delta_{0}$ after $T_0$ time slots.

\begin{algorithm}[!b]
	\setstretch{1.1}
	\renewcommand{\algorithmicrequire}{\textbf{Input:}}
	\renewcommand{\algorithmicensure}{\textbf{Output:}}
	\caption{Musical Chair Algorithm at Each server}
	\label{alg:mmab:musical-chair}
	\begin{algorithmic}[1]
		\REQUIRE $N$, $\delta_0$ (failure probability).
		\ENSURE  $f$ (external rank).
		\STATE \textbf{set} $f \leftarrow 0$,$T_0 \leftarrow N \ln \dfrac{N}{\delta_0}$; 		
		\FOR{$t=1$ to $T_0$}
			\IF{$f = 0$}
				\STATE Randomly select a sensor $i \in \mathcal{N}$
				\IF{no collision occurred}
					\STATE $f \leftarrow i$;
				\ELSE
					\STATE Select sensor $f$;
				\ENDIF
			\ENDIF
		\ENDFOR
	\end{algorithmic}
\end{algorithm}

\begin{algorithm}[!b]
	\setstretch{1.1}
	\renewcommand{\algorithmicrequire}{\textbf{Input:}}
	\renewcommand{\algorithmicensure}{\textbf{Output:}}
	\caption{Sequential Hopping Protocol at Each Server}
	\label{alg:mmab:sequential-hopping}
	\begin{algorithmic}[1]
		\REQUIRE $f$ (external rank).
		\ENSURE  $M$ (the number of servers), $R$ (rank).
		\STATE \textbf{set} $M \leftarrow 1$, $R \leftarrow 1$; 		
		\FOR{$t=1$ to $2f$}
		\STATE Select sensor $f$;
		\IF{collisions occurred}
		\STATE $R \leftarrow R+1$;
		\STATE $M \leftarrow M+1$
		\ENDIF
		\ENDFOR
		\FOR{$i=1$ to $2(N-f)$}
		\STATE Select sensor $f+i \Mod{N}$;
		\IF{collisions occurred}
		\STATE $M \leftarrow M+1$
		\ENDIF
		\ENDFOR
	\end{algorithmic}
\end{algorithm}

According to the external rank of the sensor determined in the first procedure,
the second procedure can estimate $M$ and assign different
ranks in $\left\lbrace 1,\ldots,M\right\rbrace $ to all servers using the
so-called sequential hopping protocol (Algorithm~\ref{alg:mmab:sequential-hopping}). 
During the waiting period, the server will select the sensor $f$ 
for $2f$ time slots, and then sweep through the sensor
$f+1,f+2,\ldots,N$. In this process, servers will have a collision with each
other. Then each server can calculate the number of servers $M$ by counting the number
of collisions in this process. Moreover, the rank $R$ can be obtained by
counting the collisions during the waiting period.

Subsequently, the initialization scheme will take $N \ln N/ \delta_0+2N = N \ln(e^2N / \delta_0)$ time slots. This completes the proof of Theorem~\ref{the:initialization-scheme-time-slots}
\section{Proof of Lemma \ref{pro:mmab:performance-running-consensus}}
\label{sec:mmab:proof-running-consensus}
\begin{proof}
Let $\bm{u}_{x}$ be the unit eigenvector corresponding to $\lambda_{x}$. Note that $\lambda_{1}=1$ and $\bm{u}_{1}=\frac{\bm{1}_{M}}{\sqrt{M}}$, we can rewrite \eqref{eq:mmab:samples-per-user} as
\begin{equation}
\label{eq:mmab:eigendecomposition}
\begin{aligned}
&\hat{\bm{n}}_{i}(t) =\sum_{\tau=1}^{t} \bm{S}^{t-\tau+1} \boldsymbol{p}_{i}(\tau) \\
&=\sum_{\tau=1}^{t}\sum_{x=1}^{M} \lambda_{x}^{t-\tau+1} \bm{u}_{x} \bm{u}_{x}^{\top} \boldsymbol{p}_{i}(\tau) \\
&=\frac{1}{M}\sum_{\tau=1}^{t}\bm{1}_{M} \bm{1}_{M}^{\top} \boldsymbol{p}_{i}(\tau)+\sum_{\tau=1}^{t}\sum_{x=2}^{M} \lambda_{x}^{t-\tau+1} \bm{u}_{x} \bm{u}_{x}^{\top} \boldsymbol{p}_{i}(\tau) \\
&=\overline{n}_{i}(t) \bm{1}_{M}+\sum_{\tau=1}^{t} \sum_{x=2}^{M} \lambda_{x}^{t-\tau+1} \bm{u}_{x} \bm{u}_{x}^{\top} \boldsymbol{p}_{i}(\tau)\text{.}
\end{aligned}
\end{equation}
The second equality is true since $\bm{S}$ is also a real symmetric matrix capable of using eigendecomposition. For \eqref{eq:mmab:eigendecomposition}, the $k$-th entry of the second term is bounded as follows:
\begin{equation}
\begin{aligned}
    &|\sum_{\tau=0}^{t} \sum_{x=2}^{M} \lambda_{x}^{t-\tau+1} (\bm{u}_{x} \bm{u}_{x}^{\top} \boldsymbol{p}_{i}(\tau))_k|\\ & \leq \sum_{\tau=1}^{t} \sum_{x=2}^{M}\left|\lambda_{x}^{t-\tau+1}\right|\left\|\bm{u}_{x}\right\|_{2}^{2}\left\|\boldsymbol{p}_{i}(\tau)\right\|_{2} \\
& \leq \sqrt{M} \sum_{\tau=1}^{t} \sum_{x=2}^{M}\left|\lambda_{x}^{t-\tau+1}\right| \leq \epsilon_g \text{.}
\end{aligned}
\end{equation}
Then we obtain the bound in \eqref{eq:mmab:selection-estimate-interval}. Similarly, we can rewrite \eqref{eq:mmab:reward-per-user} as
\begin{equation}
    \hat{\bm{g}}_{i}(t) =\sum_{\tau=1}^{t} P^{t-\tau+1} \boldsymbol{r}_{i}(\tau)
\end{equation}
which satisfies:
\begin{equation}
\label{eq:mmab:expected-reward-per-user}
\mathbb{E}\left[\hat{\bm{g}}_{i}(t)\right]=\mu_{i} \sum_{\tau=1}^{t} P^{t-\tau+1} \boldsymbol{p}_{i}(\tau)=\mu_{i} \hat{\bm{n}}_{i}(t) \text{.}
\end{equation}
From Sec.~\ref{sec:mmab:running_consensus}, we have $\hat{\mu} _{i}^{k}(t) = \frac{\hat{g}_{i}^{k}(t)}{\hat{n}_{i}^{k}(t)}$.
Then we can verify $E\left[ \hat{\mu} _{i}^{k}(t)\right] =\mu_i$ by
\eqref{eq:mmab:expected-reward-per-user}. 
This completes the proof of Lemma \ref{pro:mmab:performance-running-consensus}.
\end{proof}

\section{Proof of Theorem \ref{the:mmab:subopt-upper-bound}}
\label{sec:mmab:proof-subopt-upper-bound}

\begin{proof}
Denote the confidence term of UCB and LCB as 
\begin{equation}
   C_{i}^k(t) = \sqrt{ \frac{2\ln Mt}{m_i^k(t-1)}} =\sqrt{ \frac{2\ln Mt}{M\hat{n}_i^k(t-1)}}.
\end{equation}
For each server $k$, $\mathbbm{1}\left\lbrace \cdot\right\rbrace $ is the indicator function which is equal to $1$ if the condition is true and $0$ otherwise. The number of times that all the servers incorrectly select a sensor $i$ until time $T$ satisfies
\begin{equation}
\begin{aligned}
&\sum_{k=1}^{M}\tilde{m}^k_i(T) \leq M+\sum_{k=1}^{M}\sum_{t=N+1}^{T} \mathbbm{1}\left\lbrace \mathcal{E}_i^k(t)\right\rbrace  \\
\leq & W+\sum_{k=1}^{M}\sum_{t=N+1}^{T} \mathbbm{1}\left\{\mathcal{E}_i^k(t), M\overline{n}_i(t-1) \geq W\right\} \\
\leq & W+\sum_{k=1}^{M}\sum_{t=N+1}^{T}\left(\mathbbm{1}\left\{\mathcal{E}_i^k(t), \mu_{i}<\mu_{h^k(t)}, M\overline{n}_i(t-1) \geq W \right\}\right.\\
&\left.+\mathbbm{1}\left\{\mathcal{E}_i^k(t), \mu_{i}>\mu_{h^k(t)}, M\overline{n}_i(t-1) \geq W\right\}\right) \text{,}
\end{aligned}
\end{equation}
where $W$ is a positive integer that will be defined later and $\mathcal{E}_i^k(t)$ is the event that server $k$ select the sensor $i$ but $i \neq h^k(t)$ at time $t$. Denote $\mathcal{R}_{h^k(t)}^*$ as the index set $\{ 1,...,h^k(t) \}$ of the $h^k(t)$ best sensors. In the case where $\mu_{i}<\mu_{h^k(t)}$, sensor $i$ is selected implies that there exists a sensor $j \in \mathcal{R}_{h^k(t)}^*$ such that $j \notin \mathcal{R}_{h^k(t)}$. So the following inequality holds.
\begin{equation}
\begin{aligned}
&\sum_{k=1}^{M}\sum_{t=N+1}^{T}\mathbbm{1}\left\{\mathcal{E}_i^k(t), \mu_{i}<\mu_{h^k(t)}, M\overline{n}_i(t-1) \geq W\right\}\\
& \leq \sum_{k=1}^{M}\sum_{t=N+1}^{T} \mathbbm{1} \left\{   U^k_j(t) \leq  U^k_i(t)  ,M\overline{n}_i(t-1) \geq W \right\}\\
&\leq \sum_{k=1}^{M}\sum_{t=N+1}^{T} \mathbbm{1} \left\{ \min _{0<m_{j}^k(t-1)<t} U^k_j(t) \leq \max _{0 < m_i^k(t-1) <t} U^k_i(t)\right\}\\
&\leq \sum_{k=1}^{M}\sum_{t=1}^{\infty} \sum_{m_{j}^k=1}^{t-1} \sum_{m_{i}^k=1}^{t-1} \mathbbm{1}\left\{U^k_j(t) \leq U^k_i(t)\right\} \text{,}
\end{aligned}
\end{equation}
$U^k_j(t) \leq U^k_i(t)$ implies that at least one of the following inequalities must be true:
\begin{equation}
\label{eq:mmab:inequality1}
\hat{\mu}_{j}^k (t-1) \leq \mu_{j}-C_{i}^k(t) \text{,}
\end{equation}
\begin{equation}
\label{eq:mmab:inequality2}
\hat{\mu}_{i}^k(t-1)  \geq \mu_{i}+C_{i}^k(t) \text{,}
\end{equation}
\begin{equation}
\label{eq:mmab:inequality3}
\mu_{j} <\mu_{i}+2 C_{i}^k(t) \text{.}
\end{equation}
For $\delta > 0$, Hoeffding's inequality is given by 
\begin{equation}
\label{eq:mmab:Hoeffding-inequality}
P\left\{\left|\Ex[\hat{ X }_n]-\hat{ X }_n \right| \geq \delta\right\} \geq 2 e^{-2n \delta^{2}} \text{,}
\end{equation}
where $X_i \in [0,1]$,$ \hat{ X }_n = \frac{1}{n} \sum_{i=1}^n X_i $, then we can bound the probability that \eqref{eq:mmab:inequality1} and \eqref{eq:mmab:inequality2} hold using Hoeffding's inequality \eqref{eq:mmab:Hoeffding-inequality} :
\begin{equation}
P\left\{\hat{\mu}_{j}^k (t-1) \leq \mu_{j}-C_{i}^k(t) \right\} \leq e^{-4 \ln Mt} = (Mt)^{-4} \text{,}
\end{equation}
\begin{equation}
P\left\{ \hat{\mu}_{i}^k(t-1)  \geq \mu_{i}+C_{i}^k(t) \right\} \leq e^{-4 \ln Mt} = (Mt)^{-4} \text{.}
\end{equation}
The third inequality \eqref{eq:mmab:inequality3} is equivalent to
\begin{equation}
2 C_{i}^k(t)=2 \sqrt{ \frac{2\ln Mt}{M\hat{n}_i^k(t-1)}} >\mu_{j} - \mu_{i} \geq \Delta_{ \min } \text{.}
\end{equation}
\begin{equation}
\Rightarrow   \frac{8\ln Mt}{M\Delta^2_{ \min }} >\hat{n}_i^k(t-1) \geq \overline{n}_i(t-1) - \epsilon_g  \text{,}
\end{equation}
where $\Delta_{ \min }= \min \{\left| \mu_{i} - \mu_{j} \right|  \mid i,j \in \mathcal{N},\mu_{i} \neq \mu_{j} \}$.
So \eqref{eq:mmab:inequality3} does not hold if 
\begin{equation}
\overline{n}_i(t-1) >   \frac{8\ln Mt}{M\Delta^2_{ \min }}  + \epsilon_g \text{.}
\end{equation}
Therefore, when we set  $W \geq \left\lceil\frac{8 \ln MT}{\Delta^2_{ \min }}+M \epsilon_g \right\rceil$, \eqref{eq:mmab:inequality3} is false.

For the case where $\mu_i > \mu_{h^k(t)}$, we split the condition as 
\begin{equation}
\label{eq:mmab:split-condition}
\begin{aligned}
&\mathbbm{1} \left\{ \mathcal{E}_i^k(t), \mu_{i}>\mu_{h^k(t)}, M\overline{n}_i(t-1) \geq W \right\}\\
&=\mathbbm{1}\left\{\mathcal{E}_i^k(t), \mu_{i}>\mu_{h^k(t)}, M\overline{n}_i(t-1) \geq W, \mathcal{R}_{h^k(t)} = \mathcal{R}_{h^k(t)}^* \right\}\\
&+\mathbbm{1}\left\{\mathcal{E}_i^k(t), \mu_{i}>\mu_{h^k(t)}, M\overline{n}_i(t-1) \geq W, \mathcal{R}_{h^k(t)} \neq \mathcal{R}_{h^k(t)}^* \right\}\\
&\leq \mathbbm{1}\left\{ M\overline{n}_i(t-1) \geq W, L^k_i(t) \leq L^k_{h^k(t)}(t) \right\}\\
&+\mathbbm{1}\left\{M\overline{n}_i(t-1) \geq W, L^k_i(t) \leq L^k_{d}(t) \right\}
\end{aligned}
\end{equation}
for any sensor $d \in \mathcal{R}_{h^k(t)}$ but $d \notin \mathcal{R}_{h^k(t)}^*$. When sensor $i$ is selected, the two indicator functions in \eqref{eq:mmab:split-condition} can be
combined as follow in order to conclude the possibility for the case $\mu_i > \mu_{h^k(t)}$.
\begin{equation}
\begin{aligned}
&\mathbbm{1} \left\{ \mathcal{E}_i^k(t), \mu_{i}>\mu_{h^k(t)}, M\overline{n}_i(t-1) \geq W \right\}\\
&\leq \mathbbm{1} \left\{  M\overline{n}_i(t-1) \geq W, L^k_i(t) \leq L^k_a(t) \right\}
\end{aligned}
\end{equation}
for any sensor $v \notin \mathcal{R}_{h^k(t)}^* \backslash \{I_{h^k(t)}\}$. Subsequently, similar to the first case, we can have
\begin{equation}
\begin{aligned}
&\sum_{k=1}^{M}\sum_{t=N+1}^{T} \mathbbm{1} \left\{ \mathcal{E}_i^k(t), \mu_{i}>\mu_{h^k(t)}, M\overline{n}_i(t-1) \geq W \right\}\\
&\leq \sum_{k=1}^{M}\sum_{t=1}^{\infty} \sum_{m_{v}^k=1}^{t-1} \sum_{m_{i}^k=1}^{t-1} \mathbbm{1} \left\{  M\overline{n}_i(t-1) \geq W, L^k_i(t) \leq L^k_v(t) \right\}
\end{aligned}
\end{equation}
Note that $L^k_i(t) \leq L^k_v(t)$ implies that at least one of the following must be true:
\begin{equation}
\hat{\mu}_{i}^k(t-1)  \leq \mu_{i}-C_{i}^k(t) \text{,}
\end{equation}
\begin{equation}
\hat{\mu}_{v}^k(t-1)  \geq \mu_{v}+C_{v}^k(t) \text{,}
\end{equation}
\begin{equation}
\mu_{i} <\mu_{v}+2 C_{i}^{k}(t) \text{.}
\end{equation}
In the same way, we again bound the probability that the first two inequalities hold
\begin{equation}
P\left\{ \hat{\mu}_{i}^k(t-1)  \leq \mu_{i}-C_{i}^k(t) \right\} \leq (Mt)^{-4} \text{,}
\end{equation}
\begin{equation}
P\left\{ \hat{\mu}_{v}^k(t-1)  \geq \mu_{v}+C_{h}^k(t) \right\} \leq (Mt)^{-4} \text{.}
\end{equation}
As before, the third inequality never holds when we set  $W \geq \left\lceil\frac{8 \ln MT}{\Delta^2_{ \min }}+M \epsilon_g \right\rceil$.

In summary, the expected number of times that all the servers incorrectly select a sensor $i$ until time $T$ satisfies 
\begin{equation}
\begin{aligned}
&\sum_{k=1}^{M}\Ex\left[\tilde{m}^k_i(T)\right] \\ &\leq  W + \sum_{k=1}^{M}\sum_{t=1}^{\infty} \sum_{m_{j}^k=1}^{t-1} \sum_{m_{i}^k=1 }^{t-1}(P\left\{\hat{\mu}_{j} (t-1) \leq \mu_{j}-C_{i}^k(t) \right\} 
\\&+P\left\{ \hat{\mu}_{i}(t-1)  \geq \mu_{i}+C_{i}^k(t) \right\})\\
&+\sum_{k=1}^{M}\sum_{t=1}^{\infty} \sum_{m_{v}^k=1}^{t-1} \sum_{m_{i}^k=1}^{t-1}(P\left\{ \hat{\mu}_{i}^k(t-1)  \leq \mu_{i}-C_{i}^k(t)  \right\}\\
&+P\left\{ \hat{\mu}_{v}^k(t-1)  \geq \mu_{v}+C_{v}^k(t) \right\})\\
& \leq \left\lceil\frac{8 \ln MT}{\Delta^2_{ \min }}+M \epsilon_g \right\rceil +2 \sum_{k=1}^{M}\sum_{t=1}^{\infty} \sum_{m_{j}^k=1}^{t-1} \sum_{m_{i}^k=1 }^{t-1} 2(Mt)^{-4}\\
& \leq \frac{8 \ln MT}{\Delta^2_{ \min }}+M \epsilon_g+\frac{2 \pi^{2}}{3M^3}+1 \text{.}
\end{aligned}
\end{equation}
This completes the proof of Theorem \ref{the:mmab:subopt-upper-bound}.
\end{proof}


\end{document}